\documentclass[11pt]{article}
\PassOptionsToPackage{hyperfootnotes=false}{hyperref}

\usepackage[preprint]{acl}

\usepackage{times}
\usepackage{latexsym}

\usepackage[T1]{fontenc}

\usepackage[utf8]{inputenc}

\usepackage{microtype}
\microtypesetup{expansion=false}

\usepackage{graphicx}
\usepackage{caption}
\usepackage{cuted}
\usepackage{placeins}
\usepackage{tabularx}
\usepackage{array}
\usepackage{amssymb}

\usepackage{adjustbox}

\title{OmegaUse-SOP: SOP Engineering for Professional Computer Use \\ from Human Demonstrations}

\author{
  \textbf{Yixiong Xiao\textsuperscript{1}}, \textbf{Lang An\textsuperscript{1}}, 
  \textbf{Hucheng Yang\textsuperscript{2}}, \textbf{Pinxue Ma\textsuperscript{2}}, 
  \textbf{Yongquan Chen\textsuperscript{1}}, \textbf{Jingjia Cao\textsuperscript{1}} 
  \\
  \textbf{Yusai Zhao\textsuperscript{1}}, \textbf{Ting Wang\textsuperscript{2}}, 
  \textbf{Ting Liu\textsuperscript{2}}, \textbf{Siqi Bao\textsuperscript{1}}, 
  \textbf{Jingbo Zhou\textsuperscript{1}}\thanks{Project lead and corresponding author.}, \textbf{Hua Wu\textsuperscript{1}}
  \\
  \textsuperscript{1}Baidu, Inc., Beijing, China; 
  \textsuperscript{2}Ningxia Electric Power Engineering Co., Ltd., China
 \\\{xiaoyixiong, anlang, chenyongquan, caojingjia, zhaoyusai, baosiqi, zhoujingbo, wu\_hua\}@baidu.com\\
\{yanghucheng, mapinxue, wangting, liuting\}@nepdi.com.cn
}

\begin{document}
\maketitle
\begin{strip}
\centering
\includegraphics[width=0.6\textwidth]{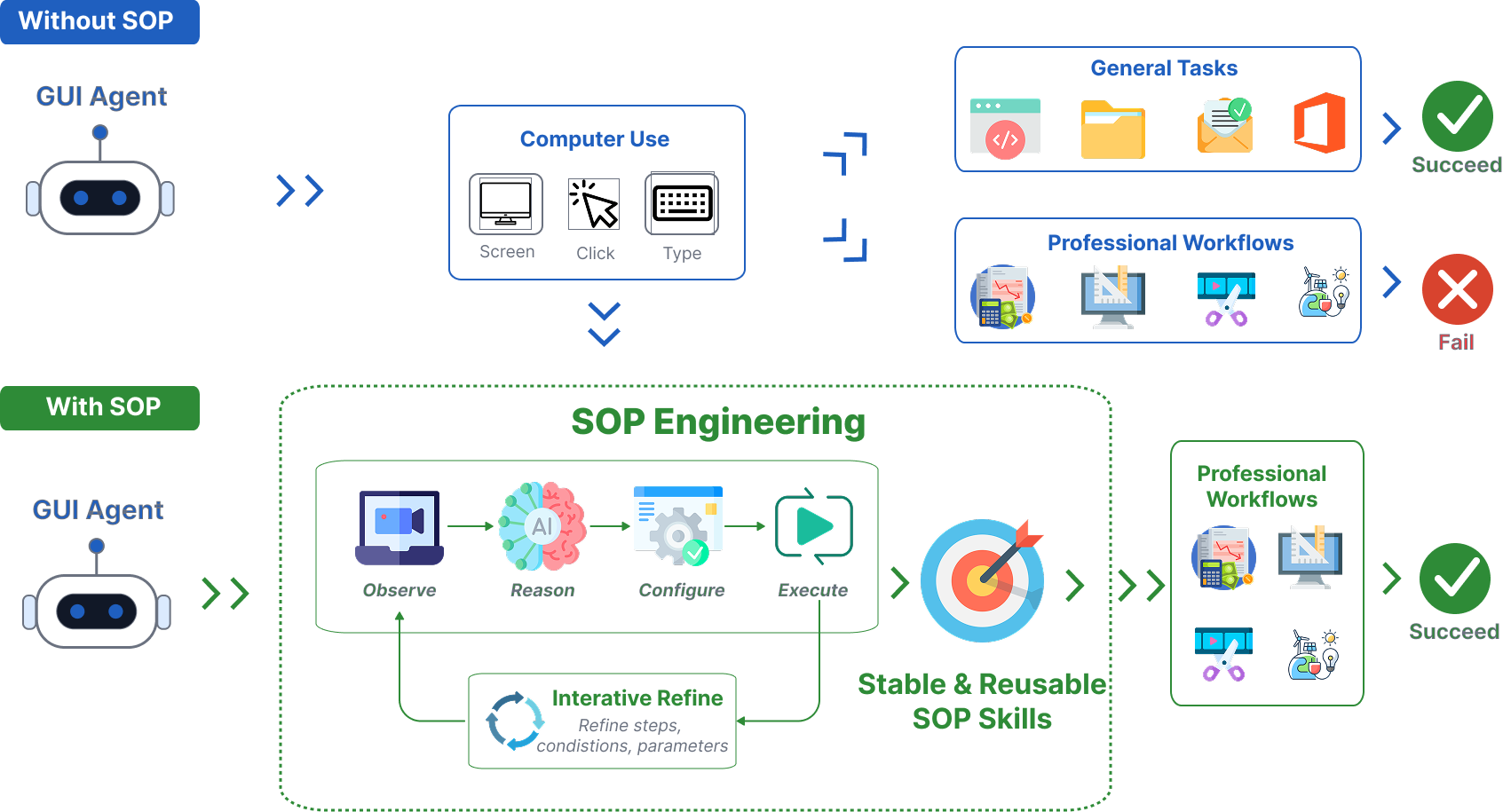}
\captionof{figure}{Comparison between GUI agents without SOP (top) and with SOP (bottom).}
\label{fig:sop-demo}
\end{strip}

\begin{abstract}

Large language models (LLMs) are increasingly evolving from conversational assistants into agents capable of operating external digital environments. Graphical user interface (GUI) agents play an important role in this transition, as many real-world workflows remain accessible only through user-facing software interfaces. However, despite recent progress on general computer-use benchmarks, domain-specific professional standard operating procedures (SOPs) remain challenging for GUI agents because they often involve implicit domain knowledge, software-specific conventions, and task-level verification requirements. We introduce \textbf{OmegaUse-SOP}, a human-in-the-loop \textbf{SOP Engineering} system for transforming human demonstrations of professional computer use into reusable SOP skills for GUI agents. Analogous to prompt engineering, SOP Engineering iteratively refines demonstrations, execution rules, and domain knowledge to convert professional SOPs into reusable GUI-agent skills. OmegaUse-SOP consists of four modules: Observe, Reason, Configure, and Execute. Together, these modules record expert operations as multimodal GUI traces, abstract low-level events into semantic step-level instructions, incorporate domain rules and task-specific parameters, and execute the resulting skills in live GUI environments through step-wise grounding, action generation, and verification. To demonstrate its effectiveness, we collaborate with a power-sector client and test OmegaUse-SOP on photovoltaic simulation workflows in PVsyst 7.2. The results suggest that OmegaUse-SOP can improve GUI-agent reliability on professional SOP tasks, highlighting a practical path toward deploying GUI agents in domain-specific professional software environments. We have open-sourced the code for OmegaUse-SOP at \url{https://github.com/baidu-frontier-research/omegause-sop}, and a demo video is available at \url{https://www.youtube.com/watch?v=E3-GCaSbDPU}.

\end{abstract}

\section{Introduction}
\begin{figure*}[t]
  \centering
  \includegraphics[width=0.75\linewidth]{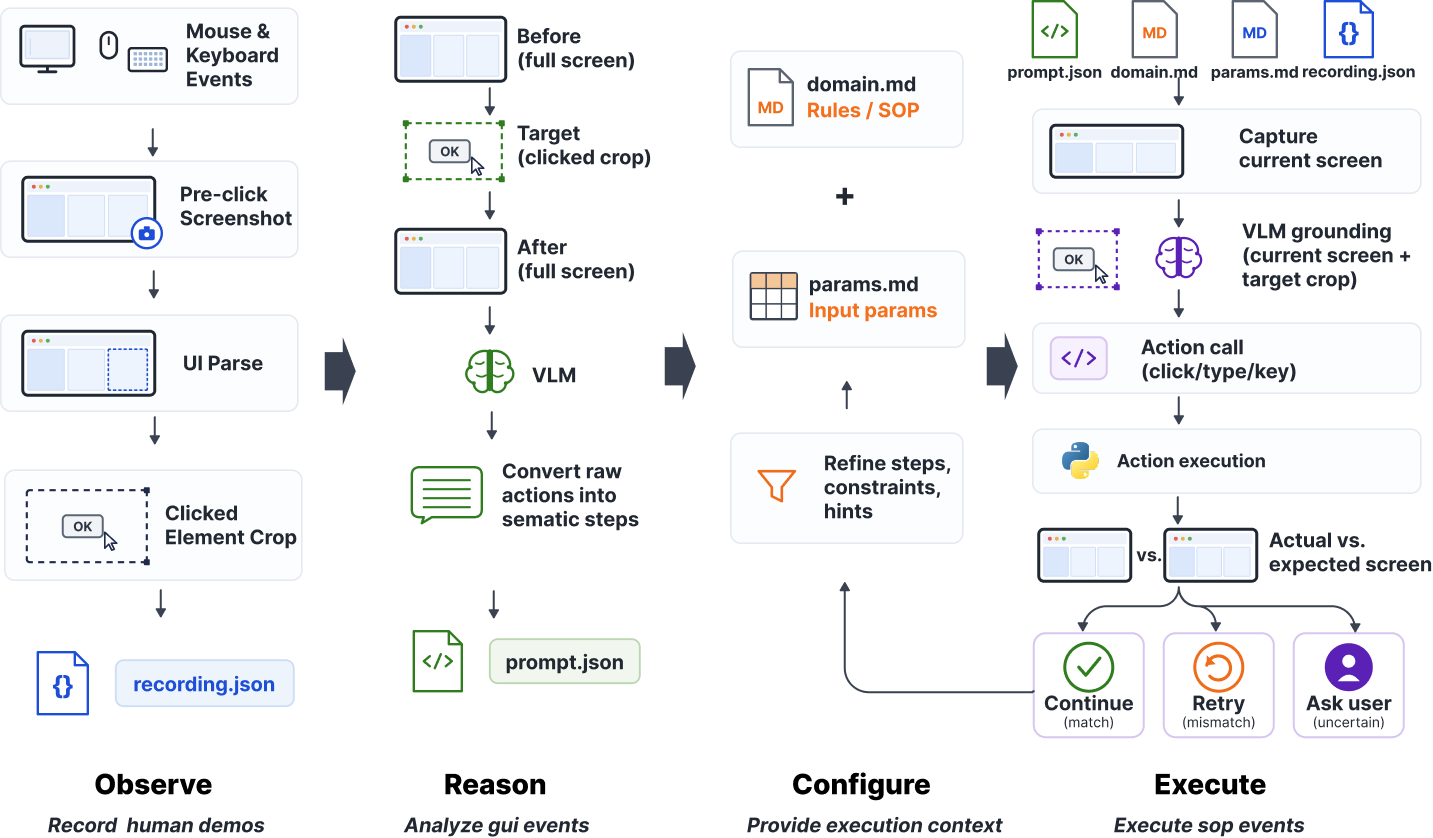}
  \caption{Framework of the OmegaUse-SOP system.}
  \label{fig:sop-eng-framework}
\end{figure*}

Large language models are increasingly developed from conversational assistants into agents that can plan, execute, and revise actions across interactive environments to help users accomplish real-world tasks. Graphical user interface (GUI) agents are an important part of this shift, as many real-world workflows remain accessible only through user-facing software interfaces rather than programmatic APIs. By perceiving screens, grounding interface elements, and acting through mouse and keyboard operations, GUI agents provide a practical interface through which LLMs can operate existing digital systems in a human-like manner  \cite{xie_osworld_2024}.

Recent work on GUI agents has advanced along two complementary directions: GUI-specialized model training and more capable agent frameworks. GUI-specialized models such as UI-TARS \cite{wang_ui-tars-2_2025}, GUI-Owl \cite{xu_mobile-agent-v35_2026}, and OmegaUse \cite{zhang_omegause_2026} improve VLMs’ GUI grounding, reasoning, and action-generation capabilities through specialized data construction, iterative model training, and end-to-end reinforcement learning. In parallel, agent frameworks such as  Agent S2 \cite{agashe_agent_2025} and UFO2 \cite{zhang_ufo2_2025}  explore task decomposition, planning, reflection, memory, multi-agent coordination, and GUI--API hybrid control, thereby improving GUI-agent execution beyond model-level capabilities alone. Benchmarks such as OSWorld~\cite{xie_osworld_2024} provide realistic desktop environments for evaluating open-ended computer-use tasks. Beyond academic research, commercial frontier systems have also demonstrated computer-use capabilities: OpenAI's Codex \cite{openai2026codexcomputeruse} and Anthropic's Claude Code \cite{anthropic2026claudecomputeruse} can interpret screen states and operate graphical interfaces through mouse and keyboard actions. Together, these developments point toward rapid progress in general-purpose computer-use agents.

However, professional computer use presents challenges that are not fully captured by general computer-use benchmarks. In many professional settings, successful execution depends on domain-specific standard operating procedures (SOPs), software-specific conventions, implicit expert knowledge, configurable task parameters, and task-level verification requirements. A professional workflow is often not merely a long sequence of GUI actions; it is a trained practice shaped by domain rules, user preferences, validation habits, and software-specific constraints. Recent benchmarks such as SOP-Bench~\cite{nandi_sop-bench_2026} and Workflow-GYM~\cite{zhu_workflow-gym_2026} on industrial and professional workflows also show that LLM agents struggle with ambiguity, branching logic, error handling, tool orchestration, workflow-stage omission, error propagation, objective drift, and insufficient software-specific knowledge. These limitations suggest that improving professional GUI agents requires not only stronger perception and action models, but also mechanisms for capturing and reusing professional procedural knowledge.

To address this problem, we introduce \textbf{OmegaUse-SOP}, a human-in-the-loop SOP engineering system for transforming human demonstrations of professional computer use into reusable SOP skills for GUI agents. OmegaUse-SOP implements the \textbf{SOP Engineering} methodology: analogous to prompt engineering, where users iteratively refine prompts with rules, constraints, examples, and formatting requirements, \textbf{SOP Engineering} iteratively refines demonstrations, execution rules, domain knowledge, and task-specific parameters until a GUI agent can reliably reproduce a professional procedure. Unlike simple coordinate replay or fixed workflow automation, OmegaUse-SOP preserves the demonstrated procedure while enriching it with semantic target descriptions, configurable parameters, domain-specific execution guidance, and step-wise verification. The system consists of four modules: Observe, Reason, Configure, and Execute. Observe records expert operations as multimodal GUI traces, including screenshots, mouse and keyboard events, timestamps, and visual interaction targets. Reason abstracts low-level events into semantic step-level instructions that describe the operated object, its visual context, and its role in the procedure. Configure allows users to incorporate domain rules and task-specific parameters, making the demonstrated procedure adaptable to new task instances. Execute applies the configured SOP skill in live GUI environments through progressive step retrieval, screen grounding, action generation, and result verification. Together, these modules turn human demonstrations into visually grounded, editable, and parameterized SOP skills for professional GUI-agent execution.

To evaluate OmegaUse-SOP, we collaborate with a power-sector client and conduct a case study on photovoltaic simulation workflows in PVsyst 7.2, a professional software package for photovoltaic system design. The study covers five representative SOP tasks, including meteorological data import, plane-orientation setting, grid-connected system setting, detailed-losses setting, and simulation execution. We evaluate both open-weight and proprietary vision-language models under two settings: direct task execution from high-level user instructions, and SOP-guided execution using OmegaUse-SOP. The results show that OmegaUse-SOP improves task completion across the evaluated GUI agents. We further conduct an ablation study showing that the Reason module is important for converting raw demonstrations into reusable SOP representations.

\section{System Design}
OmegaUse-SOP begins with a professional demonstration of an SOP in the target software, during which the Observe module records the process as interaction evidence. The Reason module then abstracts the recorded evidence into step-level semantic instructions. In the Configure module, the human user incorporates domain knowledge and task-specific parameters into the skill specification, so that the agent can adapt the demonstrated SOP to new task instances rather than merely replaying the original demonstration. Finally, the Execute module interprets the SOP in the context of the current interface, carries out each step, and verifies intermediate outcomes. Like prompt engineering, this SOP engineering process is iterative: demonstration, reasoning, configuration, and execution are repeatedly refined until the resulting skill can reliably support professional SOP execution in the target environment.

\subsection{Observe}

The Observe module is responsible for recording how a human performs a professional SOP within a desktop application. During the demonstration, it continuously monitors mouse and keyboard events while simultaneously capturing screenshots of the running system. To preserve the “see-act” order of human GUI operations, whenever an event is detected, the module saves the screenshot immediately before the event is executed, thereby retaining the interface state observed by the human before taking the corresponding action. In addition to screenshots, the Observe module records key event metadata, including the click position, event type, timestamp, and event sequence, enabling later stages to accurately reconstruct both the order and context of the demonstrated operation. Table~\ref{tab:observe-actions} summarizes the actions recorded by the Observe module. We categorize these actions into coordinate actions, where the interaction target is specified by a screen coordinate, and non-coordinate actions, where the action does not depend on a specific screen location. 

\begin{table*}
  \centering
  \begin{tabular}{ll}
    \hline
    \textbf{Category} & \textbf{Actions} \\
    \hline
    Coordinate actions
    & Left Click, Double Click, Right Click \\
    Non-coordinate actions
    & Type, Hotkey (e.g., Enter, Tab, Escape, Ctrl+C) \\
    \hline
  \end{tabular}
  \caption{\label{tab:observe-actions}
    Action categories supported by the Observe module.
  }
\end{table*}

For coordinate actions, the actual interaction target is usually the visual element located at the clicked position, such as an icon, button, menu item, or text-based control. Therefore, explicitly parsing these elements can help GUI agents better understand and reproduce human operations. The Observe module uses OmniParser \cite{lu_omniparser_2024} and PaddleOCRv5 \cite{cui_pp-ocrv5_2026} to detect UI elements in each screenshot and crops the bounding box of the element that matches the coordinate event as its visual target. For non-coordinate actions, keyboard events are handled according to their functional roles. Continuous character input is aggregated into a single text-input event, whereas special keys and shortcut combinations, such as Enter, Tab, and Escape are recorded independently.

The output of the Observe module is a structured multimodal SOP trace that contains pre-action screenshots, visual interaction targets, mouse actions, keyboard actions, timestamps, and event order. This multimodal representation provides the foundation for subsequent SOP understanding, grounding, and execution.

\subsection{Reason}

The Reason module transforms the multimodal SOP trace produced by the Observe module into step-level semantic instructions. While the Observe module records low-level execution evidence, including action types, click coordinates, and keyboard inputs, these records are difficult for a GUI agent to interpret directly. A raw coordinate indicates where the human clicked, but it does not explain which interface element was selected or how the same target should be identified on a future screen. The Reason module addresses this limitation by using a vision-language model to ground each observed action in its visual context and generate a natural-language instruction for reproducing the operation.

For coordinate actions, such as left click, double click, and right click, the performed action and its screen coordinate are already known from observation. Therefore, the goal of reasoning is not to decide what action should be taken, but to identify the semantic target of the recorded coordinate. To support this process, the model is provided with the pre-action screenshot, a visual indication of the clicked region, and the post-action screenshot when available. The pre-action screenshot shows the interface state in which the human made the selection, the clicked-region image highlights the local target, and the post-action screenshot provides evidence of the effect caused by the interaction. Based on these visual cues, the model generates a concise instruction that describes the clicked element, its relative position, and the intended interaction, enabling a future GUI agent to locate and operate on the same target rather than blindly replaying a pixel coordinate.

For non-coordinate actions, such as Type and Hotkey, the action content is already recorded by the Observe module. The Reason module does not infer a target from a clicked coordinate, but uses the screenshot before the action to explain where and why the recorded keyboard action is used. For example, if the Observe module records a text input and the Reason module identifies that the input occurs in the project-name tab, it may describe the step as ``type the project name into the project-name field'' rather than simply ``enter the recorded text''. Similarly, if the Observe module records a shortcut such as Ctrl+S, the Reason module may explain it as ``save the current project after configuration''.

Overall, the Reason module performs grounding-oriented reasoning over the observed SOP trace. It does not plan new actions or infer an alternative workflow; instead, it enriches each recorded action with semantic information about the operated object, its visual context, and its role in the procedure. The resulting semantic SOP representation bridges the gap between low-level event recording and robust agent execution, making the demonstrated procedure easier to understand, reproduce, and adapt across similar interface states.

\subsection{Configure}

The Configure module makes the semantic SOP representation editable and adaptable by incorporating user-provided domain knowledge and task-specific parameters. While the Observe and Reason modules transform a human demonstration into visually grounded step-level instructions, professional computer use often depends on information that cannot be fully inferred from screenshots or interaction traces alone. Human experts may rely on implicit domain conventions, software-specific constraints, validation habits, or preference-driven operation rules. The Configure module provides a lightweight mechanism for making such knowledge explicit before execution.

Configure introduces two types of user-editable context. The first is domain SOP guidance, which captures professional rules and software-specific execution constraints. Users can specify how certain operations should be performed, which values should be checked or replaced, whether a field should be cleared before new input, or whether a setting must be verified after modification. Such guidance may originate from existing knowledge of human experts or from post-hoc analysis of GUI-agent failures in previous trials. In this way, the Configure module supports iterative refinement of domain SOP guidance, thereby improving the agent’s success rate in subsequent executions.

The second type of context is task-specific parameters. These parameters bind the demonstrated procedure to a new task instance by identifying which recorded values should be treated as variables rather than fixed literals. This is important for professional software workflows, where the overall procedure may remain stable while inputs, identifiers, numerical values, or case-specific options change across tasks. During execution, the agent can use these parameters to replace demonstrated values with current task values while preserving the structure of the original SOP.

Together, domain SOP guidance and task-specific parameters make the configured SOP editable, reusable, and adaptable. The former captures professional execution rules, while the latter specifies which values should change across task instances. This allows the agent to apply the SOP to new professional contexts without merely replaying the original demonstration.
\begin{table*}[!t]
  \centering
  \small
  \begin{tabular}{lcccccc}
    \hline
    \textbf{Task}
    & \multicolumn{3}{c}{\textbf{w/o SOP}}
    & \multicolumn{3}{c}{\textbf{w/ SOP}} \\
    \cline{2-7}
    & \textbf{Qwen3-VL} & \textbf{GPT-5.5} & \textbf{Opus-4.7}
    & \textbf{Qwen3-VL} & \textbf{GPT-5.5} & \textbf{Opus-4.7} \\
    \hline
        Meteorological Data Importation
      & -- & -- & -- & \checkmark & \checkmark & \checkmark \\
      Plane Orientation Setting
      & -- & \checkmark & \checkmark & \checkmark & \checkmark & \checkmark \\
     Grid Connected System Setting
      & -- & \checkmark & -- & \checkmark & \checkmark & \checkmark \\
    Detailed Losses Setting
      & -- & -- & -- & \checkmark & \checkmark & \checkmark \\
    Simulation Execution
      & \checkmark & \checkmark & \checkmark & \checkmark & \checkmark & \checkmark \\
    \hline
    \textbf{Pass Rate}
      & \textbf{1/5} & \textbf{3/5} & \textbf{2/5}
      & \textbf{5/5} & \textbf{5/5} & \textbf{5/5} \\
    \hline
  \end{tabular}
  \caption{\label{tab:main-experiment}
    Performance comparison of GUI agents on PVsyst tasks with and without OmegaUse-SOP. Qwen3-VL denotes Qwen3-VL-235B-A22B-Instruct. Each entry reports the
  best outcome over three trials, with task success manually assessed
  by domain experts. }
\end{table*}

\subsection{Execute}

The Execute module applies the configured SOP to a live desktop environment. As shown in Figure~\ref{fig:sop-eng-framework}, the module leverages four types of information collected or generated throughout the OmegaUse-SOP system: the raw low-level trace recorded by the Observe module, the semantic step understanding derived by the Reason module, the domain knowledge specified in the Configure module, and the task-specific parameters specified in the Configure module. Together, these sources allow the GUI agent to execute the SOP with access to both the original demonstration evidence and the user-configured execution context.

Given that professional SOPs may contain hundreds of steps, passing all four types of information associated with every step to the GUI agent at once would impose substantial context overhead and negatively affect execution accuracy. Inspired by the skill invocation mechanism of coding agents such as Claude Code, the Execute module progressively discloses step-related SOP information during execution, i.e., only the information relevant to the current step is retrieved. For each step, the system captures the current screen and retrieves the corresponding SOP information, including the low-level trace, semantic understanding, domain knowledge, and task-specific parameters. The model then produces a computer-use action, such as a click, double click, text input, keyboard shortcut, scroll, wait, or termination action. The system parses this action and executes it through low-level mouse and keyboard control.


After each step, the system performs result verification. It compares the current screen after execution with the expected post-action screen from the original demonstration and asks the model whether the procedure should continue. The verification prompt is designed to tolerate harmless visual differences, such as cursor position, timestamps, or dynamic interface content, while detecting meaningful deviations that may indicate execution failure. To support SOP debugging and controlled execution, the Execute module further introduces a human-in-the-loop mechanism, allowing users to intervene when verification detects that the post-action interface deviates from the SOP.\footnote{The complete implementation of the
human-in-the-loop mechanism is deployed in our client's environment.
For an open-source implementation, see
\url{https://github.com/ethanyxx/co-work}.} The user can then choose to continue, retry the procedure, or stop execution.

Through this loop, OmegaUse-SOP turns a human demonstration into a visually grounded, editable, and parameterized procedure for GUI agents. The system does not replace professional judgment with a black-box policy. Instead, it provides an intermediate layer through which human professional practice can be captured, inspected, adapted, and enacted in live software environments.

\section{Experiment}
We conduct a case-study experiment to evaluate whether OmegaUse-SOP improves GUI agents’ ability to execute professional SOP tasks in PVsyst 7.2. The experiment includes five representative tasks—meteorological data importation, plane orientation setting, grid-connected system setting, detailed losses setting, and production simulation execution—which are derived from real-world workflows used by our client in the power system design domain (Table~\ref{tab:main-experiment} and~\ref{tab:task-screenshots}). Completing these tasks enables our client to simulate the annual production and system efficiency of a grid-connected PV system at a specified site under configured meteorological and system design conditions. We compare three vision-language models, including two proprietary models, GPT-5.5 and Opus-4.7, and one open-weight model, Qwen3-VL-235B-A22B-Instruct, under two settings: a baseline setting, where the GUI agent directly performs each task based only on the user instruction, and an SOP Engineering setting, where the agent follows an SOP skill generated from human demonstration, semantic abstraction, and user configuration.

Table~\ref{tab:main-experiment} shows that OmegaUse-SOP improves task completion across all three models. In the baseline setting, the agents achieve different levels of success: Qwen3-VL completes 1/5 tasks, GPT-5.5 completes 3/5 tasks, and Opus-4.7 completes 2/5 tasks. The failures are especially pronounced in tasks that require professional procedural knowledge. For example, detailed losses setting requires configuring thermal behavior, ohmic losses, ageing, and other loss-related parameters across multiple sub-panels, which demands substantial domain understanding. In contrast, with OmegaUse-SOP, all three models complete five tasks, reaching a pass rate of 5/5. This consistent improvement across the evaluated models suggests that OmegaUse-SOP can make GUI-agent execution more reliable on professional PVsyst workflows.

\begin{table}[t]
  \centering
  \small
  \begin{tabular}{p{0.45\columnwidth}cc}
    \hline
    \textbf{Task} & \textbf{w/o Reason} & \textbf{w/ Reason} \\
    \hline
    Meteorological Data Importation
      & -- & \checkmark \\
    Plane Orientation Setting
      & \checkmark & \checkmark \\
    Grid Connected System Setting
      & -- & \checkmark \\
    Detailed Losses Setting
      & -- & \checkmark \\
    Simulation Execution
      & \checkmark & \checkmark \\
    \hline
    \textbf{Pass Rate}
      & \textbf{2/5}
      & \textbf{5/5} \\
    \hline
  \end{tabular}
  \caption{\label{tab:ablation-study}
    Ablation study results on Qwen3-VL-235B-A22B-Instruct, comparing task completion with and without the Reason module. Each entry reports the
  best outcome over three trials, with task success manually assessed
  by domain experts.
  }
\end{table}

A key design in the OmegaUse-SOP system is the Reason module, which helps the agent interpret a human demonstration before execution. To examine its importance, we conduct an ablation study on Qwen3-VL-235B-A22B-Instruct by comparing task completion with and without the Reason module. Without Reason, the agent directly uses the low-level trace recorded from the human demonstration; with Reason, the demonstration is first abstracted into step-level semantic instructions. As shown in Table~\ref{tab:ablation-study}, removing the Reason module reduces the pass rate from 5/5 to 2/5. The agent only completes plane orientation setting and simulation execution without Reason, while failing on meteorological data importation, grid-connected system setting, and detailed losses setting. This result suggests that the Reason module is important for converting human demonstrations into reusable SOPs for professional GUI-agent execution.
\section{Conclusion}

We demonstrate how professional procedural knowledge can be captured and reused by GUI agents through OmegaUse-SOP, a human-in-the-loop SOP engineering system. The system records expert GUI operations, abstracts them into semantic step-level instructions, incorporates domain guidance and task-specific parameters, and executes the resulting skills in GUI environments. We evaluate OmegaUse-SOP through a case study on photovoltaic simulation workflows in PVsyst 7.2, using tasks derived from real-world client workflows in the power-system design domain. The results show that OmegaUse-SOP improves task completion across both open-weight and proprietary VLMs, and that semantic reasoning over demonstrations is important for converting raw interaction traces into reusable SOP representations. Overall, OmegaUse-SOP provides a practical approach to capturing professional procedural knowledge and improving GUI-agent reliability in domain-specific professional software environments.

\section*{Acknowledgments}
This research was supported in part by the National Natural Science Foundation of China under Grant No.~92370204.
\bibliography{sop_eng}

\clearpage
\onecolumn
\appendix

\section{PVsyst Task Screenshots}
\label{app:pvsyst-task-screenshots}

\begin{center}
\normalsize
\setlength{\tabcolsep}{5pt}
\renewcommand{\arraystretch}{1.15}

\begin{adjustbox}{
  max width=\textwidth,
  max totalheight=0.90\textheight,
  keepaspectratio
}
\begin{tabular}{
  >{\raggedright\arraybackslash}m{0.22\textwidth}
  >{\centering\arraybackslash}m{0.72\textwidth}
}
  \hline
  \textbf{\large Task} & \textbf{\large Screenshot} \\
  \hline

  \textbf{Meteorological Data Importation}
  &
  \includegraphics[width=0.95\linewidth]{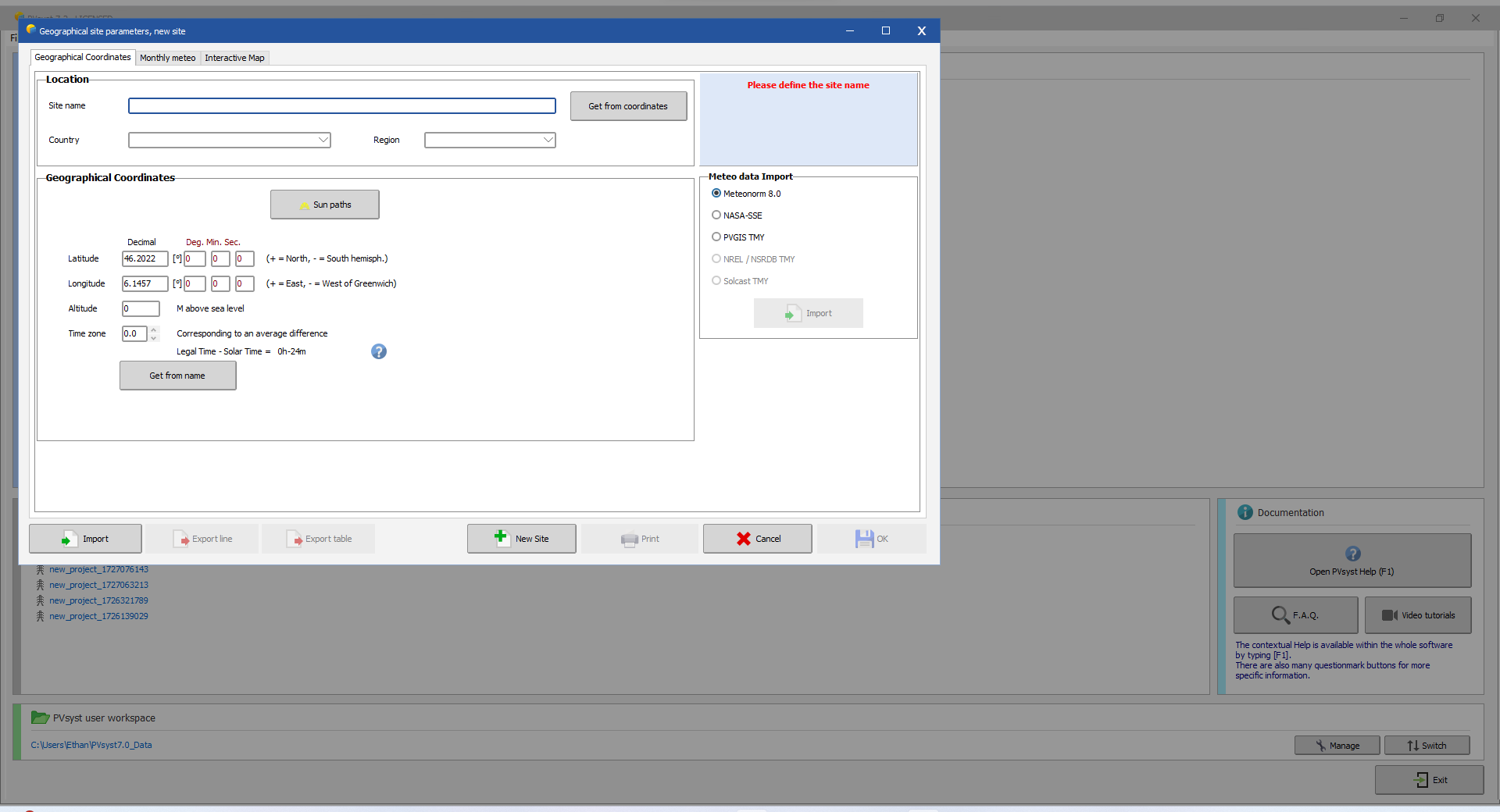}
  \\

  \textbf{Plane Orientation Setting}
  &
  \includegraphics[width=0.95\linewidth]{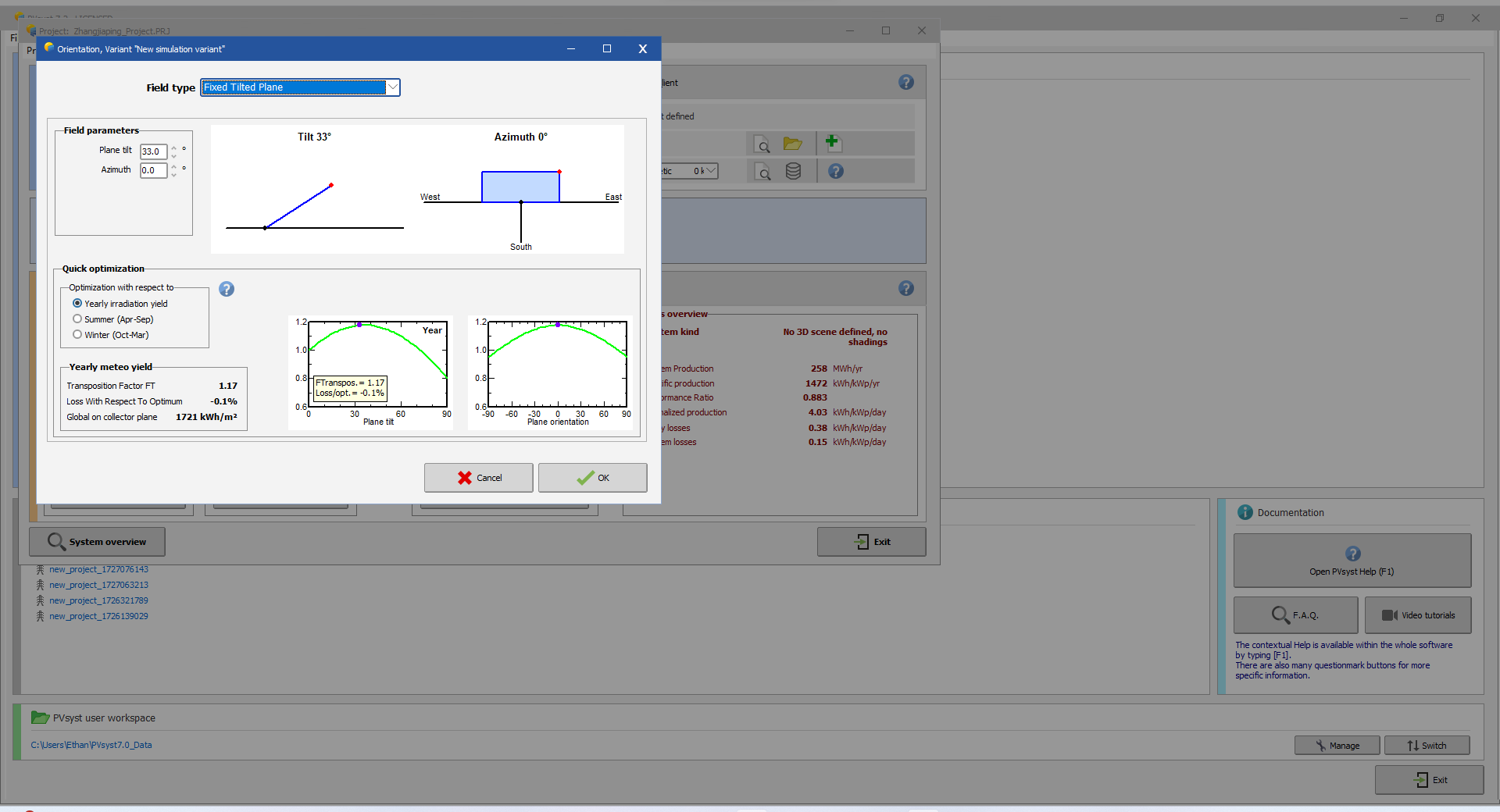}
  \\

  \textbf{Grid Connected System Setting}
  &
  \includegraphics[width=0.95\linewidth]{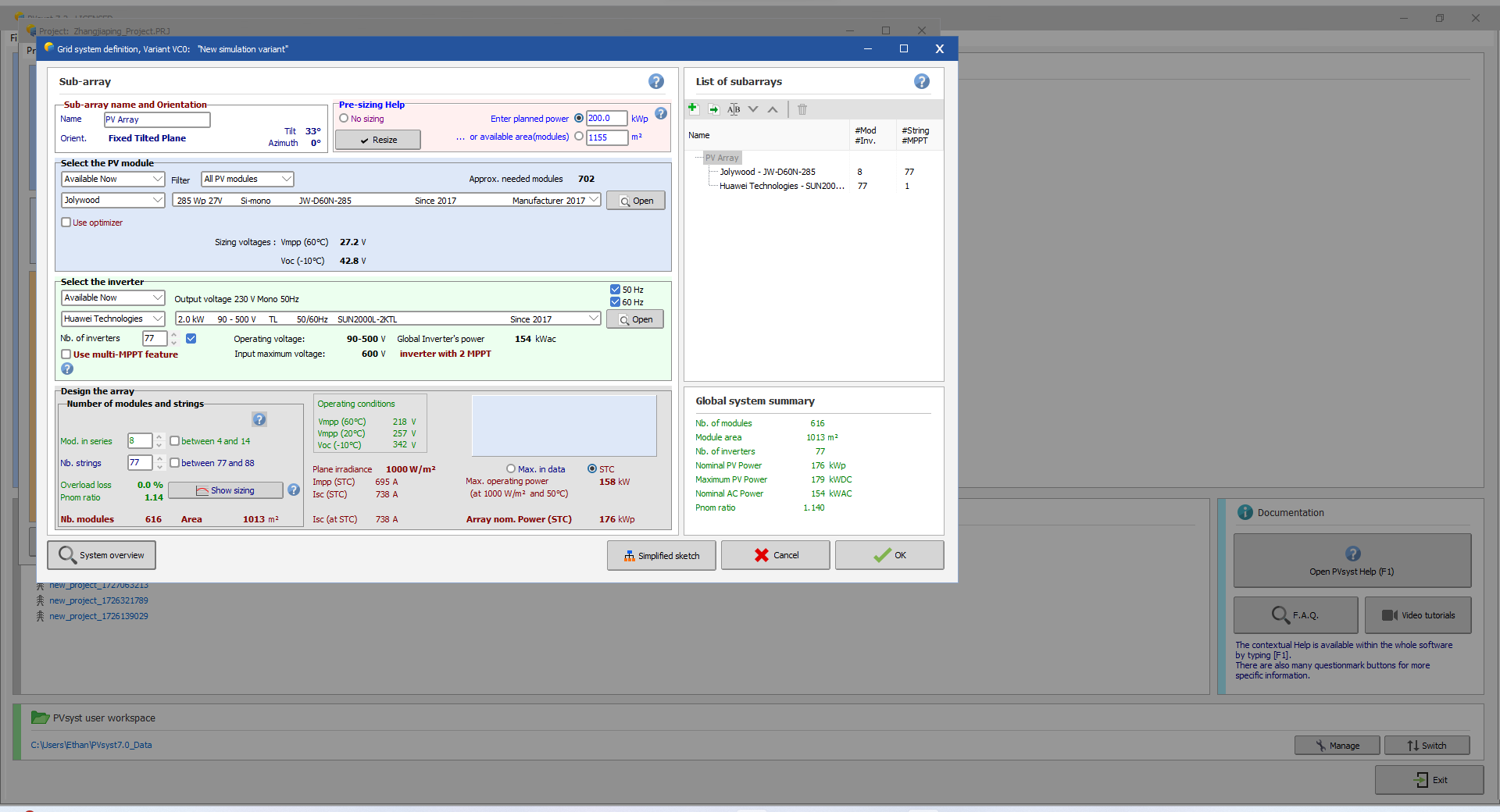}
  \\

  \textbf{Detailed Losses Setting}
  &
  \includegraphics[width=0.95\linewidth]{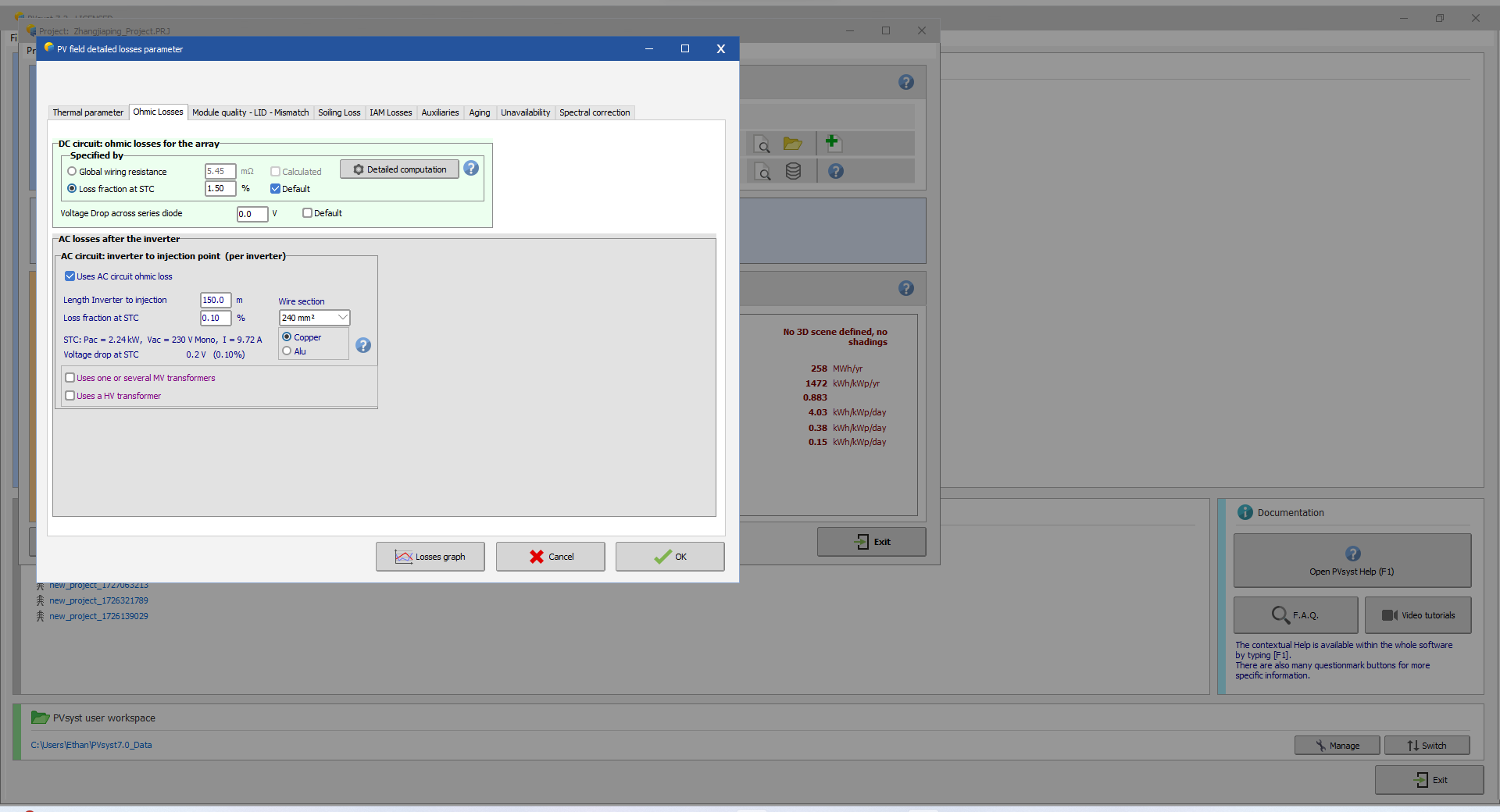}
  \\

  \textbf{Simulation Execution}
  &
  \includegraphics[width=0.95\linewidth]{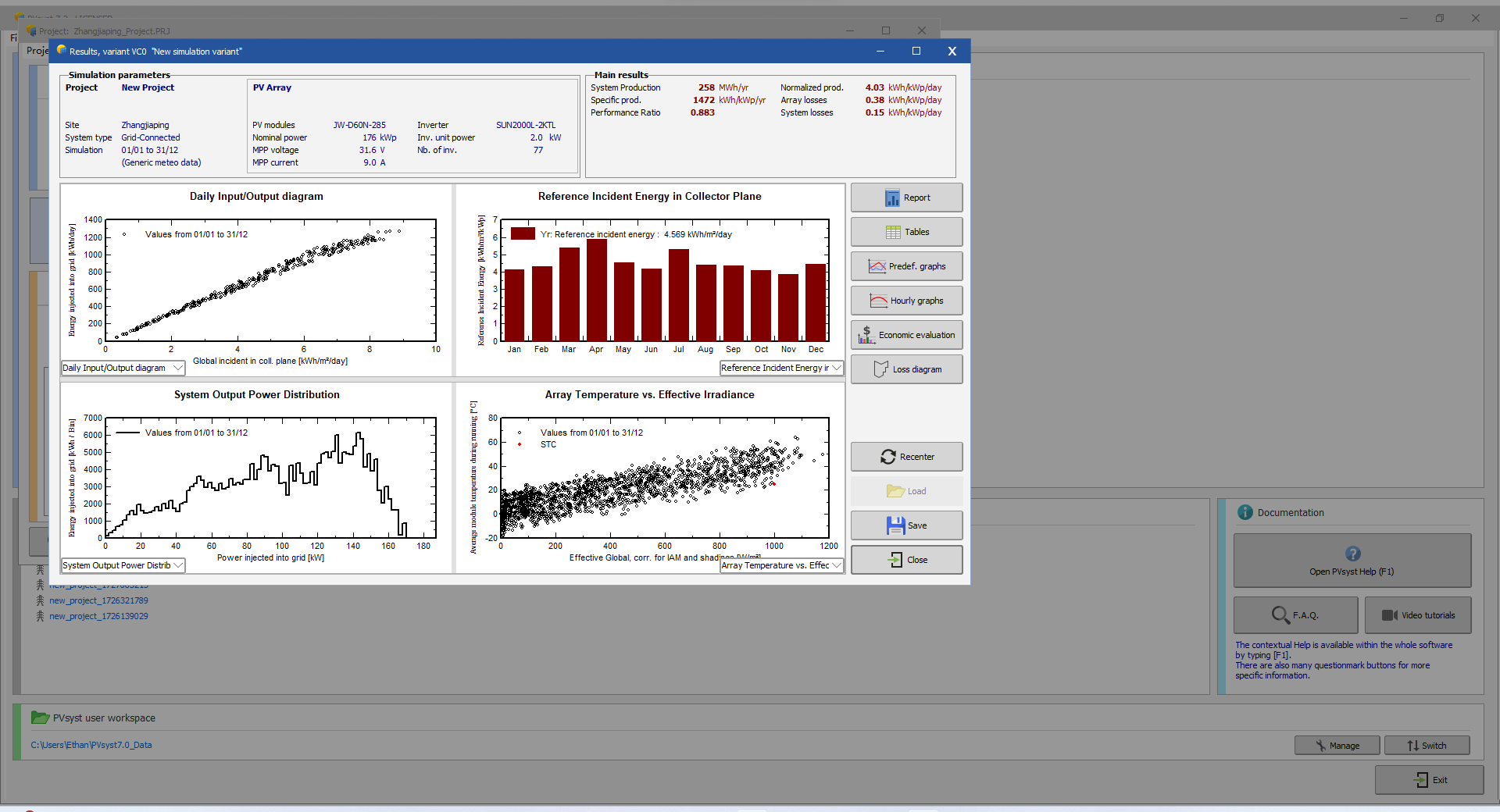}
  \\

  \hline
\end{tabular}
\end{adjustbox}

\captionof{table}{Representative PVsyst panels associated with the five GUI tasks used in the main experiment.}
\label{tab:task-screenshots}
\end{center}
\section{Illustrative Examples of Domain SOP Guidance}
\label{app:domain-sop-guidance-examples}

\begin{figure}[htbp]
  \centering
  \includegraphics[width=0.90\textwidth]
  {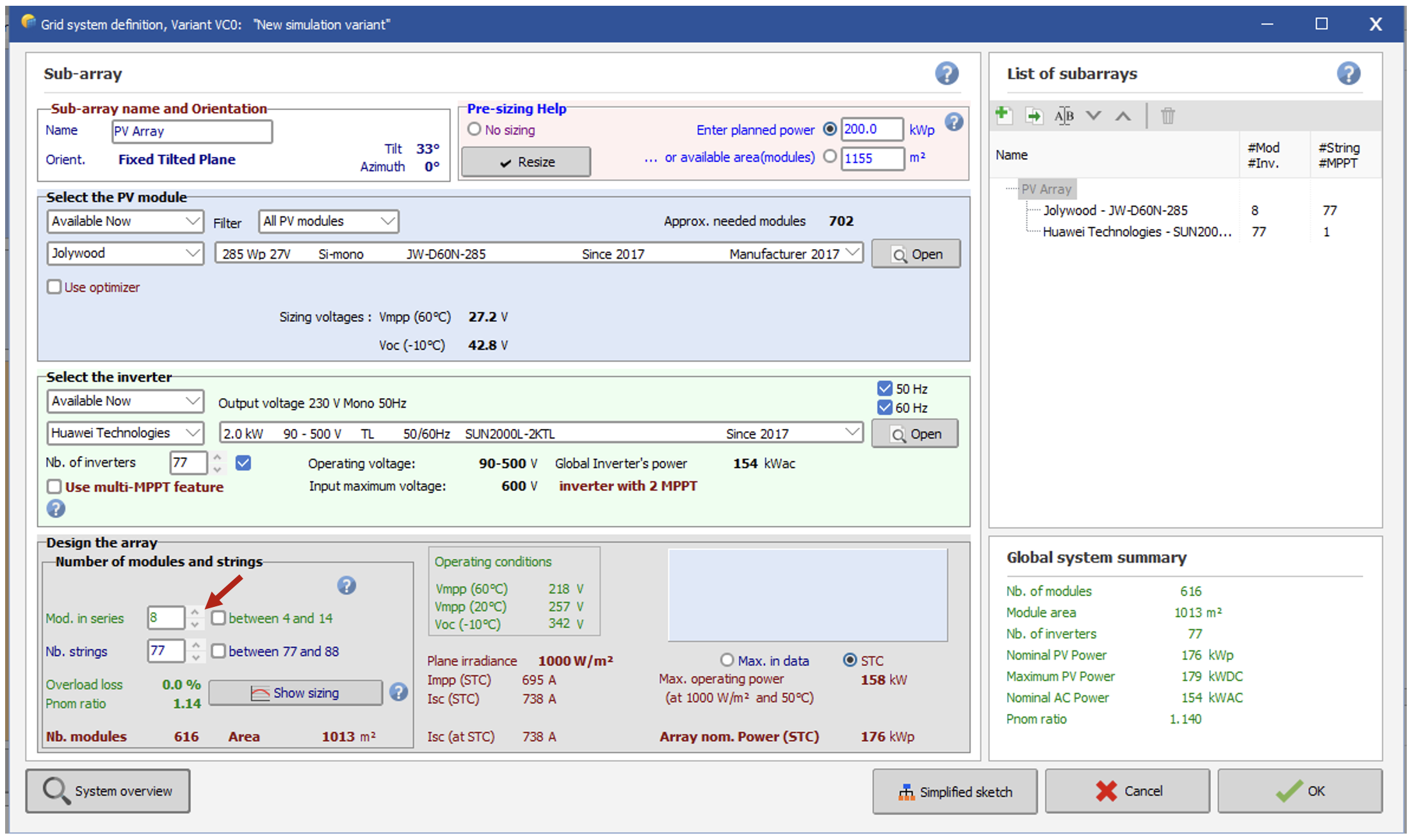}
  \caption{The PVsyst Grid Connected System Setting page, with the
  control for changing the number of modules in series marked by the
  red arrow.}
  \label{fig:expert-sop-series-modules}
\end{figure}

\begin{figure}[htbp]
  \centering
  \includegraphics[width=0.76\textwidth]
  {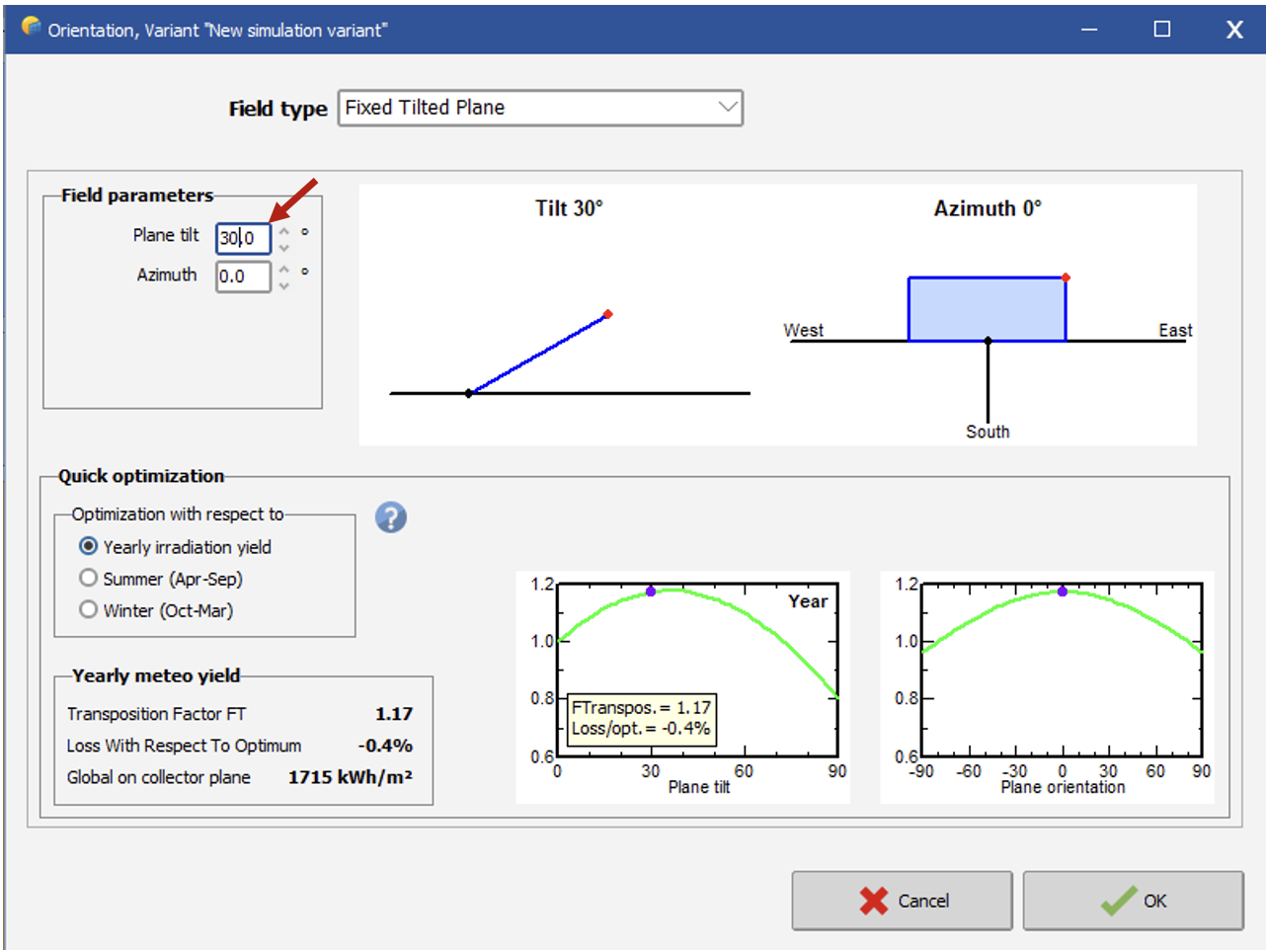}
  \caption{The PVsyst Plane Orientation Setting page, with the
  plane-tilt input field marked by the red arrow.}
  \label{fig:expert-sop-plane-tilt}
\end{figure}

Figures~\ref{fig:expert-sop-series-modules} and~
\ref{fig:expert-sop-plane-tilt} show the PVsyst interfaces associated
with two pieces of domain SOP guidance added by a domain expert after
observing model errors during trials. 

\paragraph{Example 1: Operating a Spinner Control.}
As shown in Figure~\ref{fig:expert-sop-series-modules}, during trials,
the domain expert observed that the model sometimes selected an arrow
button inconsistent with the requested direction of adjustment when
changing the number of modules in series. The expert therefore added
the following guidance:

\begin{quote}
\textit{When modifying System--Number of Modules in Series, click the
down-arrow button to decrease the value and the up-arrow button to
increase it.}
\end{quote}

\paragraph{Example 2: Prioritizing User-Specified Parameter Values.}
As shown in Figure~\ref{fig:expert-sop-plane-tilt}, during trials, the
domain expert observed that the model sometimes retained a value from
the operation instructions or historical inputs instead of applying
the parameter value explicitly specified by the user. The expert
therefore added the following guidance:

\begin{quote}
\textit{When the user specifies values for certain parameters, such as
setting the plane tilt to $35^{\circ}$, prioritize the user-specified
values over those in the operation instructions or historical inputs.}
\end{quote}

\end{document}